\documentclass[12pt]{iopart}
\usepackage{iopams}  
\usepackage{epsfig}
\usepackage{graphicx}

\begin{document}

\title[Initial State Granularity]{Constraining the initial state granularity
with bulk observables in Au+Au collisions at $\sqrt{s_{\rm NN}}=200$ GeV}

\author{Hannah Petersen${}^1$, Christopher Coleman-Smith${}^{1,2}$, Steffen A.
Bass${}^1$ and Robert Wolpert${}^2$\\[.4cm]}

\address{${}^1$~Department of Physics, Duke University, Durham, North Carolina
27708-0305, United States\\
${}^2$~Department of Statistical Science, Duke University, Durham, North
Carolina 27708-0251, United States}

\ead{hp52@phy.duke.edu}

\begin{abstract}
In this paper we conduct a systematic study of the  granularity of the 
initial state of hot and dense QCD matter produced in ultra-relativistic
heavy-ion collisions and its
influence on bulk observables like particle yields, $m_T$ spectra and elliptic
flow. For our investigation we use a hybrid transport
model, based on  (3+1)d hydrodynamics and a microscopic Boltzmann transport
approach. The initial conditions are generated by a non-equilibrium hadronic
transport approach and the size of their fluctuations can be adjusted by
defining
a Gaussian smoothing parameter $\sigma$. 
The dependence of the hydrodynamic evolution on the choices of 
   $\sigma$ and $t_{start}$ is explored by means of a Gaussian emulator.
 To generate particle yields and elliptic flow that are compatible with 
experimental data the initial state parameters are constrained to be $\sigma=1$
fm and $t_{\rm start}=0.5$ fm.  
In addition, the influence of changes in the equation of state is studied and
the results of our event-by-event calculations are compared to a calculation
with averaged initial conditions.  We conclude that even though the initial
state parameters can be constrained by yields and elliptic flow, the granularity
needs to be constrained by other correlation and fluctuation observables. 
\end{abstract}



% PACS Relativistic heavy-ion collisions, Global features in relativistic heavy
%ion collisions, Monte Carlo simulations (including hadron and parton cascades
%nd string breaking models),  Hydrodynamic models 

\maketitle

\section[Introduction]{Introduction}
\label{intro}

The study of the properties of strongly interacting matter at very high
temperatures
and/or densities is the goal of the experimental heavy ion program  
at the Relativistic Heavy Ion Collider (RHIC) at Brookhaven National
Laboratory (BNL) and at the Large Hadron Collider (LHC) at the
European Organization for Nuclear Research (CERN) in Geneva. 
In order to extract  the relevant QCD properties, such as  the
viscosity and other transport coefficients, from the final state particle
distributions measured by the experiments, effective dynamical 
approaches are needed that connect theoretical input with experimental data. 

So called hybrid approaches that apply ideal relativistic fluid dynamics for
the hot and dense stage of the collision, when the produced matter is close to
thermal equilibrium, and use microscopic 
transport approaches for the initial and/or final
non-equilibrium stages have proven to be the most successful approaches for the
description of relativistic heavy-ion collisions
\cite{Bass:2000ib,Teaney:2001av,Hirano:2005xf,Nonaka:2006yn,Petersen:2008dd,
Werner:2010aa}. 
Unfortunately, as the models gain
complexity and sophistication, the number of parameters used to encode the
relevant physics increases. All of these parameters need to be constrained
or determined  before conclusions about the properties of hot and dense
QCD matter can be drawn -- this constitutes a non-trivial multi-parameter
optimization
problem.

The {\em Models and Data Analysis Initiative} (MADAI) 
collaboration has been formed with the aim of developing novel techniques
based on statistical science to address multiparameter estimation problems 
one encounters in the application of complex models to the extraction of 
knowledge from large multi-dimensional experimental data sets.
Here, we will focus on ultra-relativistic heavy-ion collisions, in particular
on 
the problem of constraining the initial conditions of such collisions via
a comparison of bulk observables calculated in our model approach to data. 

Recently, there has been a rising interest in 
quantifying the initial state fluctuations in ultra-relativistic heavy-ion
collisions due to their importance for multi-particle correlation measurements
in particular regarding elliptic flow, triangular flow and the so-called
``ridge'' correlations
\cite{Agakishiev:2010ur,Alver:2010dn,Schenke:2010rr,Petersen:2010cw,Qin:2010pf,
Mocsy:2010um}.

In this paper a systematic study of the initial state granularity and its
influence on bulk observables such as particle yields, $m_T$ spectra and
elliptic
flow is presented. In Section \ref{hybrid} the employed hybrid transport
approach is
introduced and in Section \ref{emulator} the statistical methods are
discussed. In Section \ref{ic_tstartsig} we describe how particle yields and
elliptic flow can be used to constrain the initial state parameters. The
influence of changes in the equation of state and a comparison to averaged
initial conditions is contained in Section \ref{de_ave}. In the last Section
\ref{sum}, the conclusions from this systematic study are summarized. 

\section[Model Description]{The Hybrid Approach}
\label{hybrid}

The present hybrid model used to simulate the dynamics of Au+Au collisions at 
$\sqrt{s_{\rm NN}}=200~$GeV is based on UrQMD
\cite{Bass:1998ca,Bleicher:1999xi} with an intermediate ideal hydrodynamic
evolution for the hot and dense stage of the reaction \cite{Petersen:2008dd}. 
UrQMD is a string/hadronic transport model which simulates multiple
interactions of ingoing and newly produced particles, the excitation and
fragmentation of color strings \cite{NilssonAlmqvist:1986rx,Sjostrand:1993yb}
and the formation and decay of hadronic resonances. To mimic
experimental conditions as realistically as possible the non-equilibrium
dynamics in the initial and the final state are taken into account on an
event-by-event basis (see e.g. discussions in
\cite{Bleicher:1998wd,Grassi:2005pm,Andrade:2005tx,Andrade:2006yh,Tavares:2007mu,Andrade:2008xh}).

In UrQMD, the incoming nuclei are initialized according to Woods-Saxon profiles
and the initial nucleon-nucleon scatterings and non-equilibrium dynamics proceed
according to the Boltzmann equation. After the two nuclei have passed through
each other local
thermal equilibrium is assumed to make the transition to the ideal hydrodynamic
description. The time chosen for the mapping of the particle degrees of
freedom to the thermodynamic fields is called $t_{\rm start}$. This is one of the
parameters that we need to adjust in this calculation (default is $t_{\rm
start}=0.5$ fm). The point particles are represented by three-dimensional
Gaussian distributions in the following way

\begin{equation}
\epsilon(x,y,z)=\left(\frac{1}{2\pi}\right)^{\frac{3}{2}}\frac{\gamma_z}{
\sigma^3} E_p \exp{-\frac{(x-x_p)^2+(y-y_p)^2+(\gamma_z(z-z_p))^2}{2\sigma^2}}
\end{equation}
to obtain energy, momentum and net baryon density distributions that are smooth
enough for the hydrodynamic evolution. Here, $\epsilon$ is the energy density at
position $(x,y,z)$ that a particle with energy $E_p$ at position $(x_p,y_p,z_p)$
contributes. The Gaussians are Lorentz contracted in z-direction by $\gamma_z$
to account for the large longitudinal velocities. Only the matter at midrapidity
$|y|<2$ is assumed to be locally equilibrated, whereas the other hadrons are
treated in the hadronic cascade. The second parameter that influences the
smearing and therefore the granularity of the initial condition is the width of
the Gaussian $\sigma$. This parameter will be varied between $\sigma=0.8$ fm,
the lower limit to keep the hydrodynamic code numerically stable and
$\sigma=2$ fm which leads to very smooth profiles as shown in
\cite{Petersen:2010di}. The default choice is $\sigma=1$ fm which corresponds to
the typical size of a nucleon. 

In this manner, the initial conditions for the hydrodynamic calculation include
fluctuations from the early non-equilibrium dynamics such as finite velocity
profiles and peaks in the energy density because of fluctuations in the energy
deposition. Starting from these (single event) initial conditions a full (3+1)
dimensional
ideal hydrodynamic evolution is performed using the SHASTA algorithm
\cite{Rischke:1995ir,Rischke:1995mt}. The hydrodynamic evolution is stopped if a
certain transition criterion is fulfilled. In \cite{Petersen:2009mz} we have
explored a freeze-out procedure to account
for the large time dilatation that occurs for fluid elements at large
rapidities. To mimic an iso-$\tau$ hypersurface we freeze out full transverse
slices, of thickness $\Delta z = 0.2 $fm, whenever all cells of that slice
fulfill the freeze-out criterion ($\epsilon \approx 713~ {\rm
MeV/fm^3}$). On these slices particles are produced
according to the Cooper-Frye formula and the hadronic rescattering and resonance
decays are taken into account in the hadronic cascade (UrQMD).

The event-by-event calculation provides the full final phase-space distribution
of the hadrons that are also measured in experiments and therefore allows for
detailed comparisons of many observables  at the same time. In the present study
we concentrate on the question how the initial state granularity can be
constrained by bulk observables. In the following section, we
shall demonstrate how a more sophisticated statistical analysis helps to
get a good handle on this multi-parameter fit problem.  

Other important parameters that influence the
results are finite shear and bulk viscosity, the freeze-out criterion and the
equation of state that is employed for the calculation. Some of these parameters
are not implemented in our current model (e.g. the shear and bulk viscosities)
and therefore an extensive study
of these additional parameter dependencies is left to a future
publication. 

\section[Emulator]{Gaussian Process Emulators for Computer Models}
\label{emulator}

A thorough exploration of the dependence of UrQMD+hydrodynamics predictions of transverse
mass spectra for midrapidity pions and kaons on such model parameters as
$\sigma$ and $t_{\textrm{\tiny start}}$ could entail thousands of code
evaluations taking each $\sim$ 3 CPU-hours. As an alternative, we construct a
model
\emph{emulator} \cite {OHag:2006, Oakl:OHag:2002}--- a statistical model of
UrQMD+hydrodynamics which, after ``training'' with the model results
at a designed finite set of locations in its parameter space, can
predict rapidly what the hybrid approach predictions \emph {would} be at other
untried parameter vectors, with an attendant measure of uncertainty.  Unlike
simpler interpolation schemes, which typically provide only a single
estimated value at each point in the space, the emulator generates an entire
joint probability distribution representing what is known about the unobserved
computer model outputs, supporting the generation of predictions of arbitrary
functions of the computer model outputs with measures of their uncertainty.
Numerical implementations of the emulator models are so fast that it is
entirely practical to simulate thousands or even millions of predicted model
outputs in minutes of computer time.  This makes possible a broad range of
interesting analysis of the output of computer codes which would otherwise
require an unacceptably large computational effort (see \cite{Oakl:OHag:2004,
Baya:Berg:Paul:etal:2007, Higd:Gatt:etal:2008}, for example).

\begin{figure}[ht]
 \includegraphics[width=0.5\textwidth]{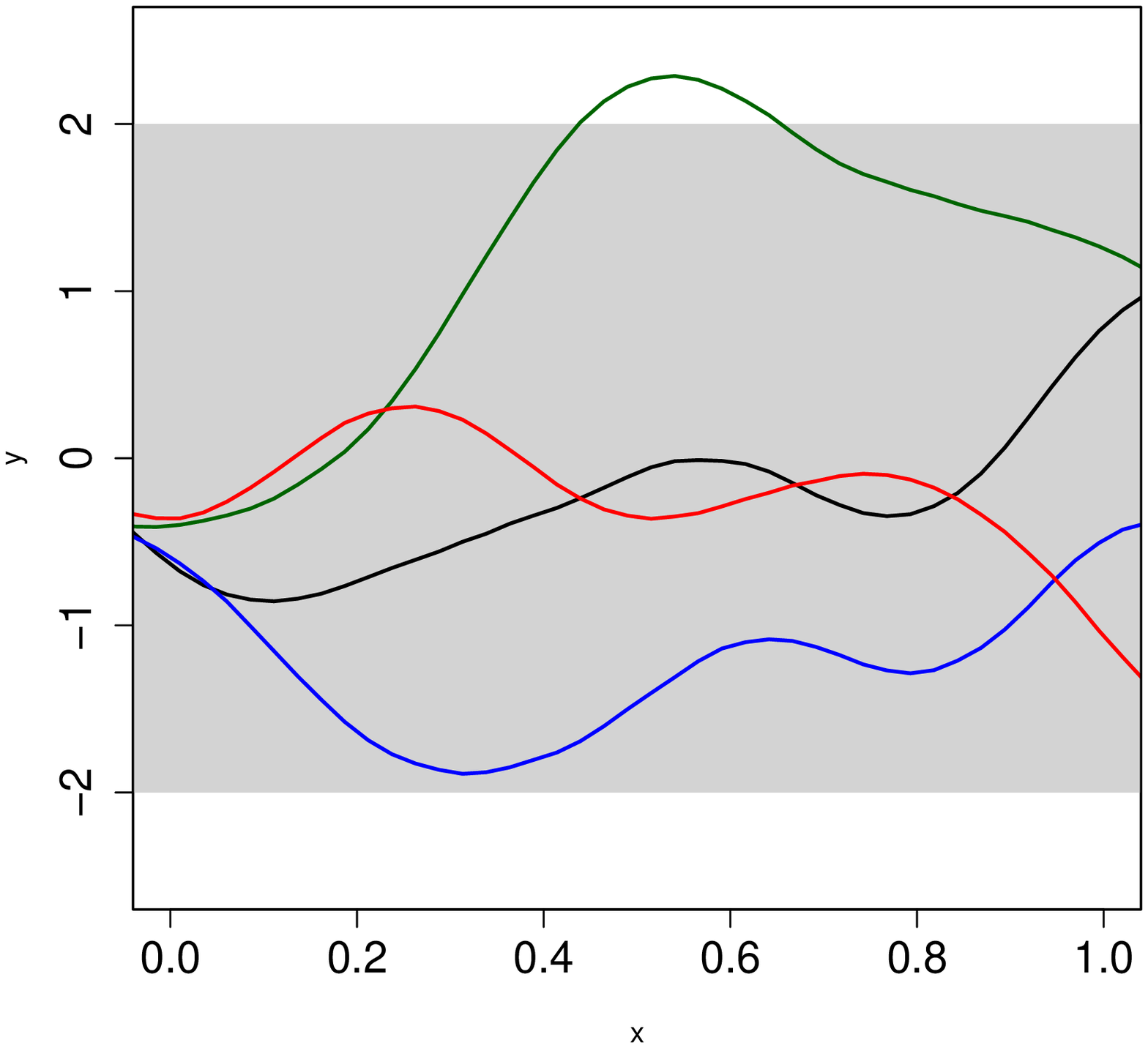} 
\includegraphics[width=0.5\textwidth]{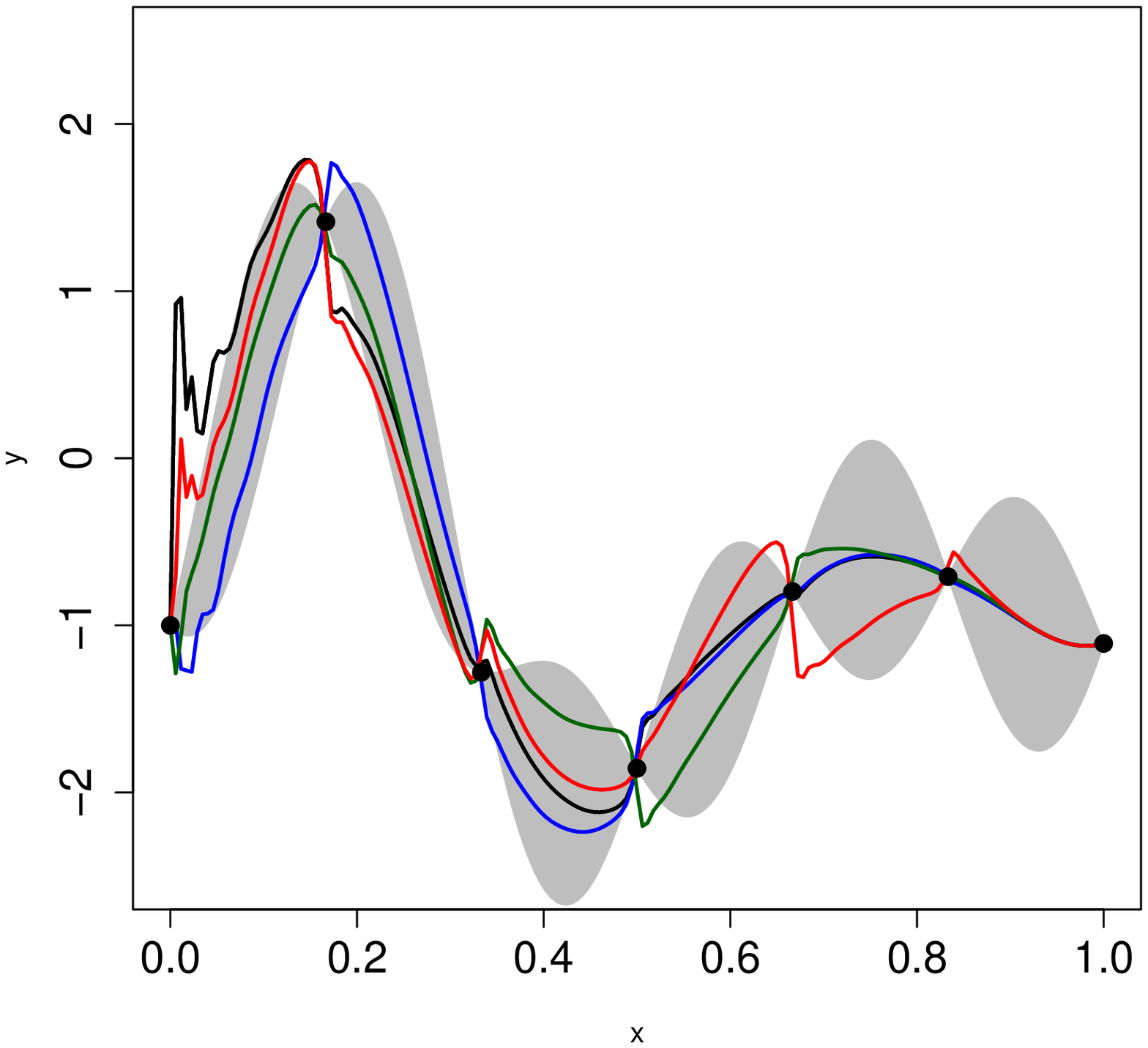}
\caption{Left: unconditioned draws from a Gaussian process. Right: draws from the same
process after conditioning on 7 training points (black circles) from a simple
model. The grey band is a 95\% confidence interval. Note how the uncertainty
grows away from the training points.}
 \label{fig-gp-example}
\end{figure}

For our emulator the computer model's output at all possible input vectors
$\{\theta\}$ is modeled as a Gaussian stochastic process or random field,
indexed by $\theta$, expressing initial uncertainty about possible model
results; several random draws from such a distribution are illustrated for a
one-dimensional parameter space in Fig. \ref
{fig-gp-example}(left).  Emulator training is accomplished by evaluating the
\emph
{conditional} probability distribution of model outputs at all input vectors
$\{\theta\}$, given observations of the actual computer model output at a
finite collection $\mathcal{D}$ of selected \emph {design} input vectors 
$\{\theta_i \in \mathcal{D} \}$.  
The emulator reproduces the model output perfectly at the
design points, and offers predictive distributions for model output at other
points which are more (or less) variable for points that are closer to (or
further from) the design points, respectively.  Although other distributions
could be used for emulators, Gaussian processes (or GPs) are particularly
convenient because the required conditional distributions can be computed
very efficiently with simple matrix algebra.

\begin{figure}[ht]\centering
 \begin{tabular}{ l c r }
 \includegraphics[width=0.25\textwidth]{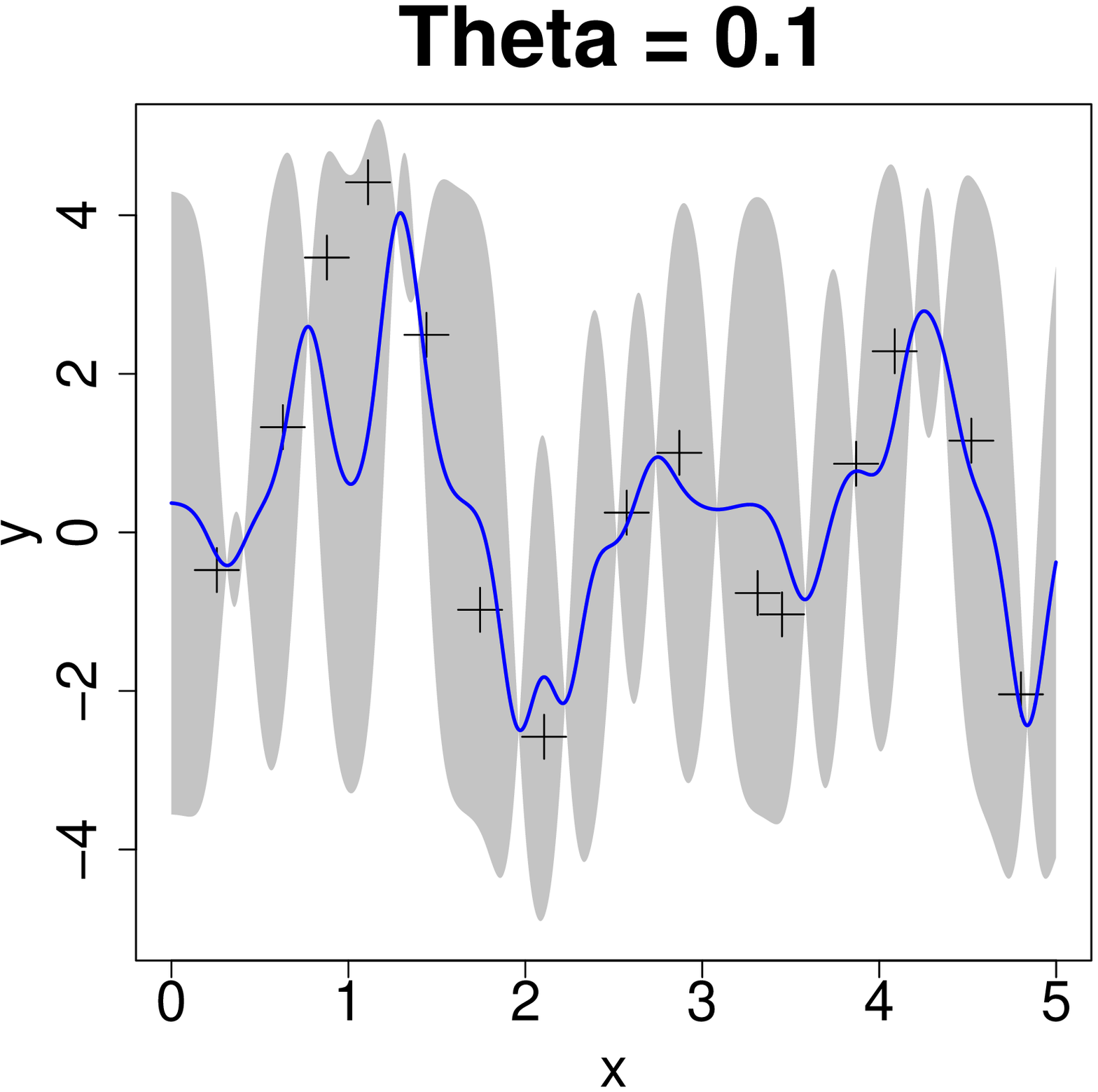} &
 \includegraphics[width=0.25\textwidth]{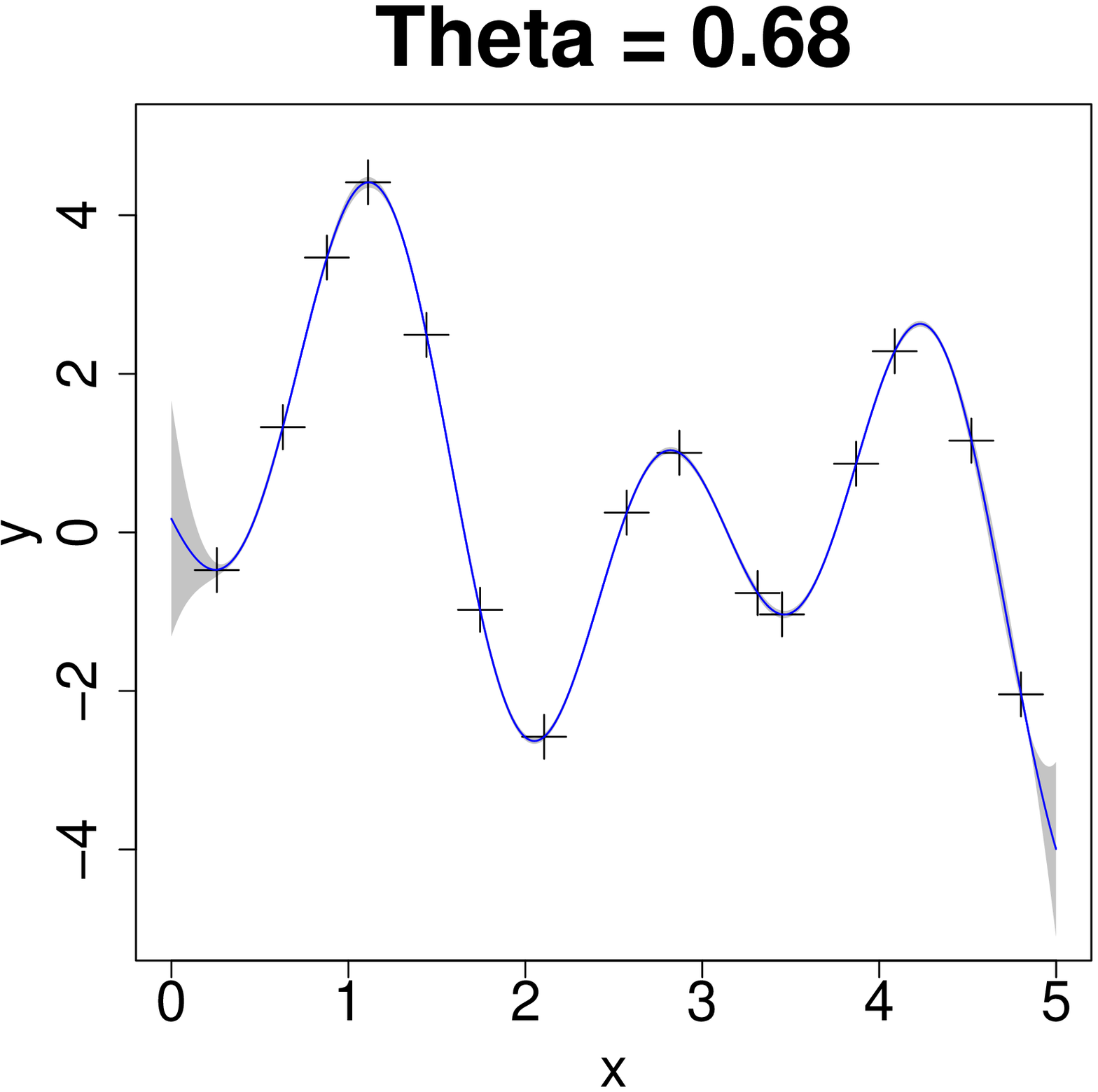}  &
 \includegraphics[width=0.25\textwidth]{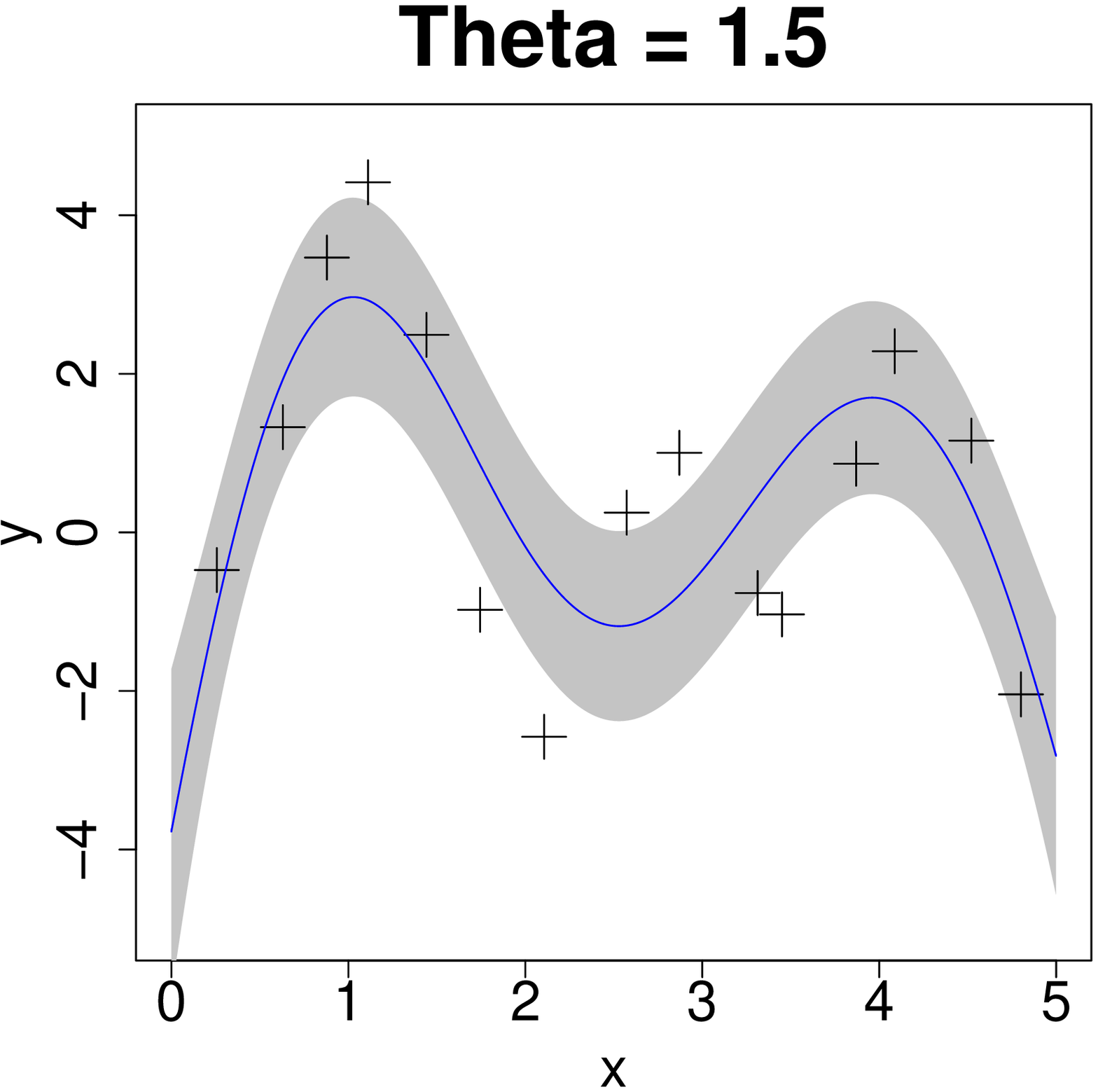}
 \end{tabular}
 \caption{Varying the correlation length $\theta$ changes the
   emulator dramatically.  Solid blue lines represent emulator means, while
   pointwise $95\%$ (pointwise) predictive intervals are shown in grey.
   Left shows over-fitting with $\theta$ too small; right shows
   over-smoothing with $\theta$ too large; middle shows optimal value of
   $\theta=0.68$, chosen by likelihood maximization.}\label{fig-gp-theta}
\end{figure}

The detailed behaviour of the emulator is determined by the mean and
covariance structure chosen for the GP and by the number and location of the
training points.  After removing simple linear effects with multiple
regression and a possible nonlinear transformation of the computer model's
inputs and outputs, if necessary, the computer model may be assumed to be
locally smooth (so small changes in $\theta$ lead to small changes in the
output) and homogeneous (so the magnitude of predictive uncertainty is
relatively constant).  Under these assumptions the conventional mean-zero
isotropic Gaussian random fields used in geostatistics \cite {Chil:Delf:1999,
Cres:1993} work well; we used the commonly-recommended power exponential
family \cite {Kenn:OHag:2000, Sack:Welc:Mitc:Wynn:1989}, with power close to
its upper limit of two to ensure smoothness, and with correlation lengths fit
to the data using maximum likelihood methods (emulator dependence on this
parameter is illustrated in Fig. \ref{fig-gp-theta}).  A modest number of
training points, perhaps 10--15 points per dimension in the parameter space,
evenly distributed over the parameter range of interest will suffice.  We used a Latin
hypercube design, a space-filling approach which has proven very successful
for all-purpose designs of computer experiment runs because it can ``fill''
the design space with very few points (see \cite {Sack:Welc:Mitc:Wynn:1989}
or \cite[Chap.~5] {Sant:Will:Notz:2003}).

To emulate \emph {functional} output, when the computer model output for each
input vector $\theta$ is not a single numerical quantity but a function of
one or more variables, such as rapidity distributions or particle spectra which
vary as a function of centrality, beam energy, etc..., we expand those functional outputs in an orthogonal
basis (the most efficient choice is to use Principal Components), then
construct independent Gaussian emulators as above for each component
separately.  This is only marginally more computationally demanding than the
emulation of univariate computer model output.

In summary: emulators are computationally-trivial statistical approximations
to functions that open exciting new doors for the statistical exploration of
computationally-expensive models.

\section{Initial State Variation}
\label{ic_tstartsig}

\begin{figure}[h]
\includegraphics[width=0.5\textwidth]{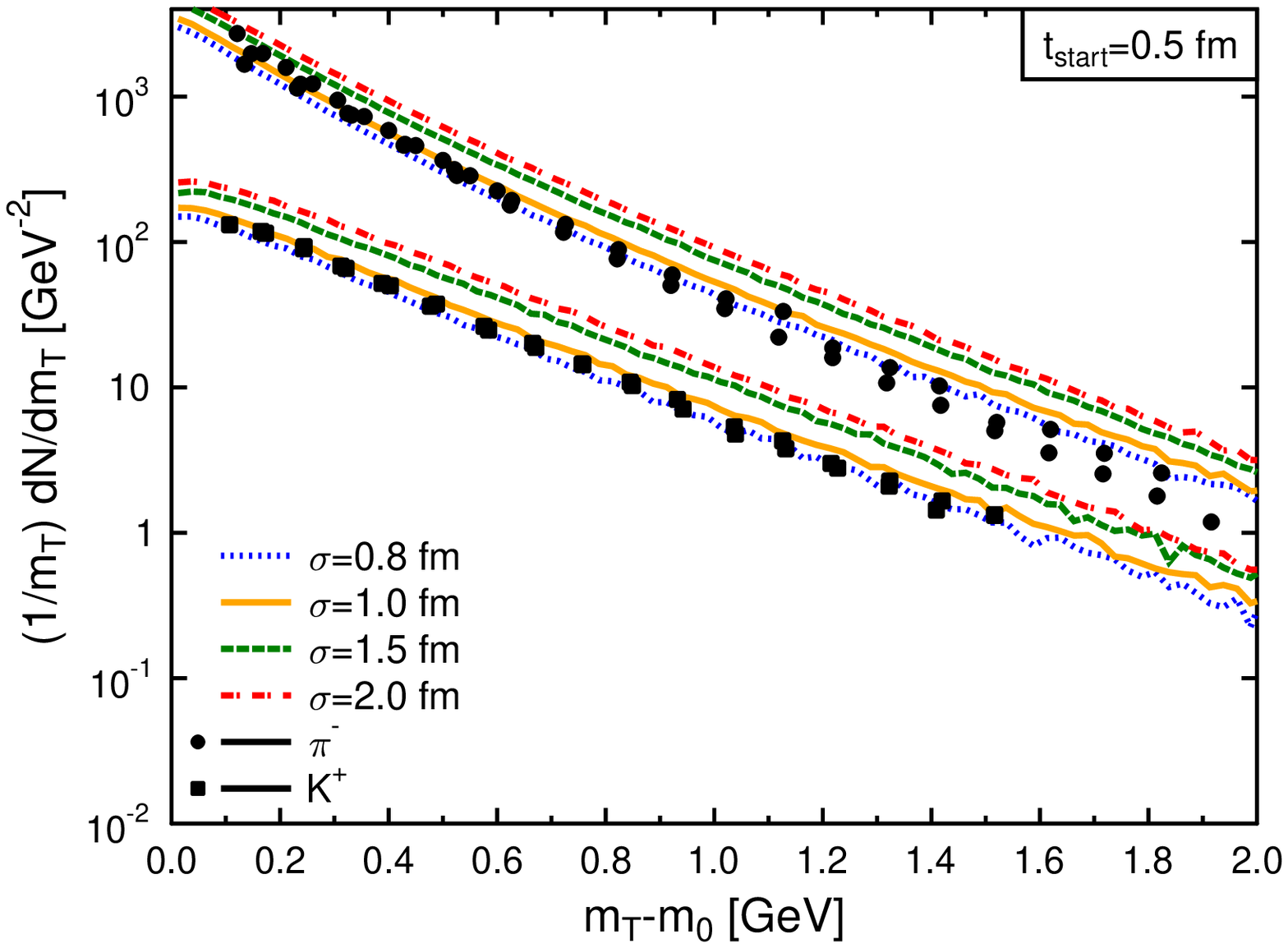}
\includegraphics[width=0.5\textwidth]{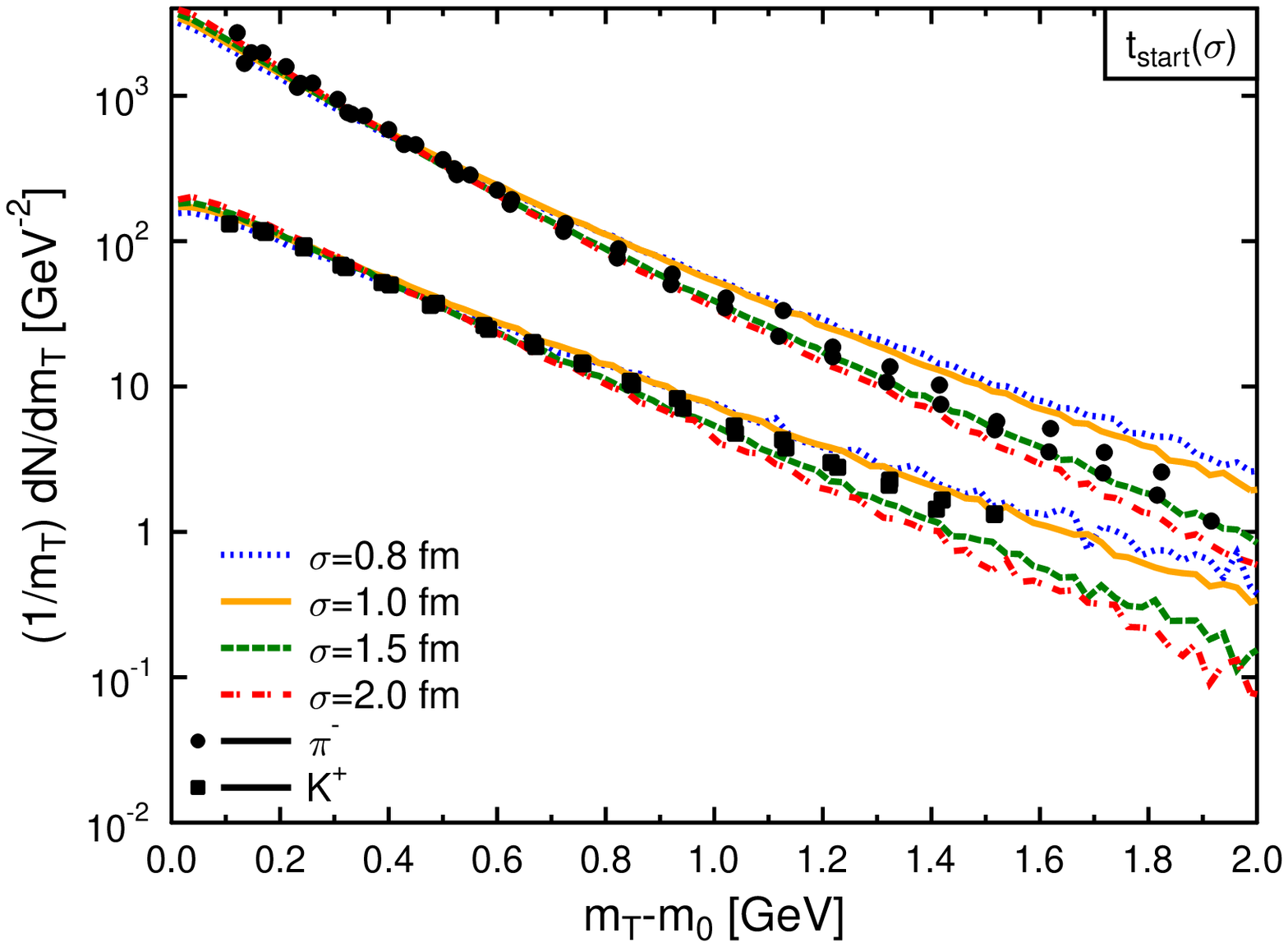}
\caption[Transverse mass spectra for $\sigma$ and $t_{\rm start}$
variation]{Transverse mass spectra for $\pi^-$ and $K^+$ at midrapidity
($|y|<0.5$) in central ($b<3.4$ fm) Au+Au collisions at $\sqrt{s_{\rm NN}}=200$
GeV from the hybrid approach for different values of $\sigma$ with a fixed
starting time (left) and a varied starting time (right) represented by lines in
comparison to experimental data that is shown as symbols
\cite{Adams:2003xp,Adler:2003cb,Arsene:2005mr}. \label{fig_dndmt_sig} }
\end{figure}

Let us now apply the emulator based on a Gaussian regression process to
determine the initial state parameters from particle yields and spectra. By
varying the Gaussian width that has been introduced in Section \ref{hybrid} the
initial state granularity can be changed while keeping everything else constant.
Fig. \ref{fig_dndmt_sig} (left) shows the result for transverse momentum spectra
for pions and kaons in central Au+Au collisions at the highest RHIC energy. The
slope of the spectra stays constant while the overall height which corresponds
to the respective particle yield increases with increasing $\sigma$. This can be
traced back to an increase in the total entropy of the initial state that is
larger, if the particles are smeared out over a larger phase space volume. 

From earlier studies in \cite{Petersen:2008dd} at lower energies it is known
that the yields are also affected by the choice of the starting time. There
is the possibility to compensate a change in $\sigma$ by adjusting 
$t_{\rm start}$ at the same time. Please note, that the starting time is here
not only a way to normalize the initial state distributions as in hydrodynamic
calculations with smooth initial conditions, but it is also the
transition time from the hadronic transport approach to ideal hydrodynamics. By
using the emulator to explore the two-dimensional parameter space in $\sigma$
and 
$t_{\rm start}$, we can find parameter combinations that lead to the same pion
yield/entropy in the system. The following parameter pairs can be identified
and are listed in table~\ref{tab_sigtstart}.

\begin{table}[h]
\begin{center}
\renewcommand{\arraystretch}{1.3}
\begin{tabular}{|l|c|c|c|c|}
\hline 
$\sigma$ [fm] & 0.8 & 1.0 & 1.5 &2.0\\
\hline
$t_{\rm start}$ [fm] &0.3 & 0.5 &1.2 &1.8 \\
\hline
\end{tabular}
\caption{\label{tab_sigtstart} Combinations of $\sigma$ and $t_{\rm start}$ that
lead to the same pion yield in central ($b<3.4$ fm) Au+Au collisions at
$\sqrt{s_{\rm NN}}=200$ GeV. These values have been used in all the calculations
that are dubbed with '$t_{\rm start}$ varied'.}
\end{center}
\end{table}

\begin{figure}[h]
\centerline{\includegraphics[width=0.6\textwidth]{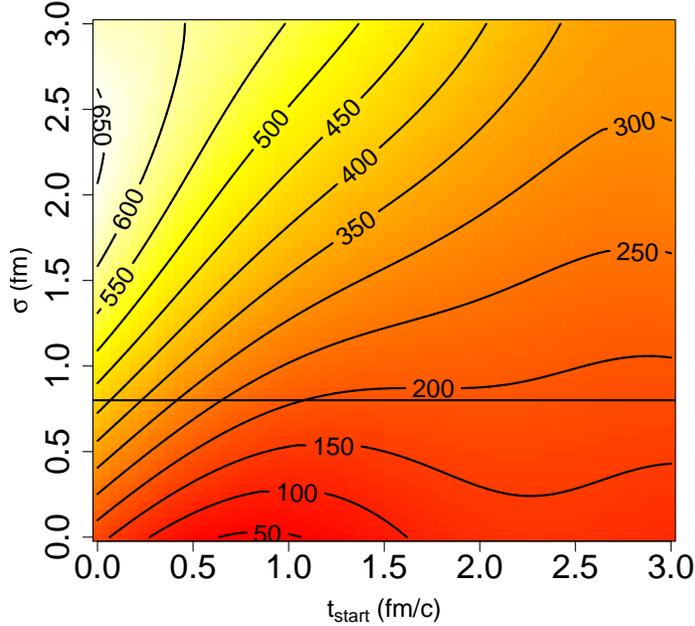}}
\caption[Emulated pion numbers]{Emulated number of pions at midrapidity
($|y|<0.5$) for central ($b<3.4$ fm) Au+Au collisions at  $\sqrt{s_{\rm
NN}}=200$ GeV in the two-dimensional parameter space of $\sigma$ and $t_{\rm
start}$. \label{fig_emu_pions} }
\end{figure}

\begin{figure}[h]
\includegraphics[width=0.5\textwidth]{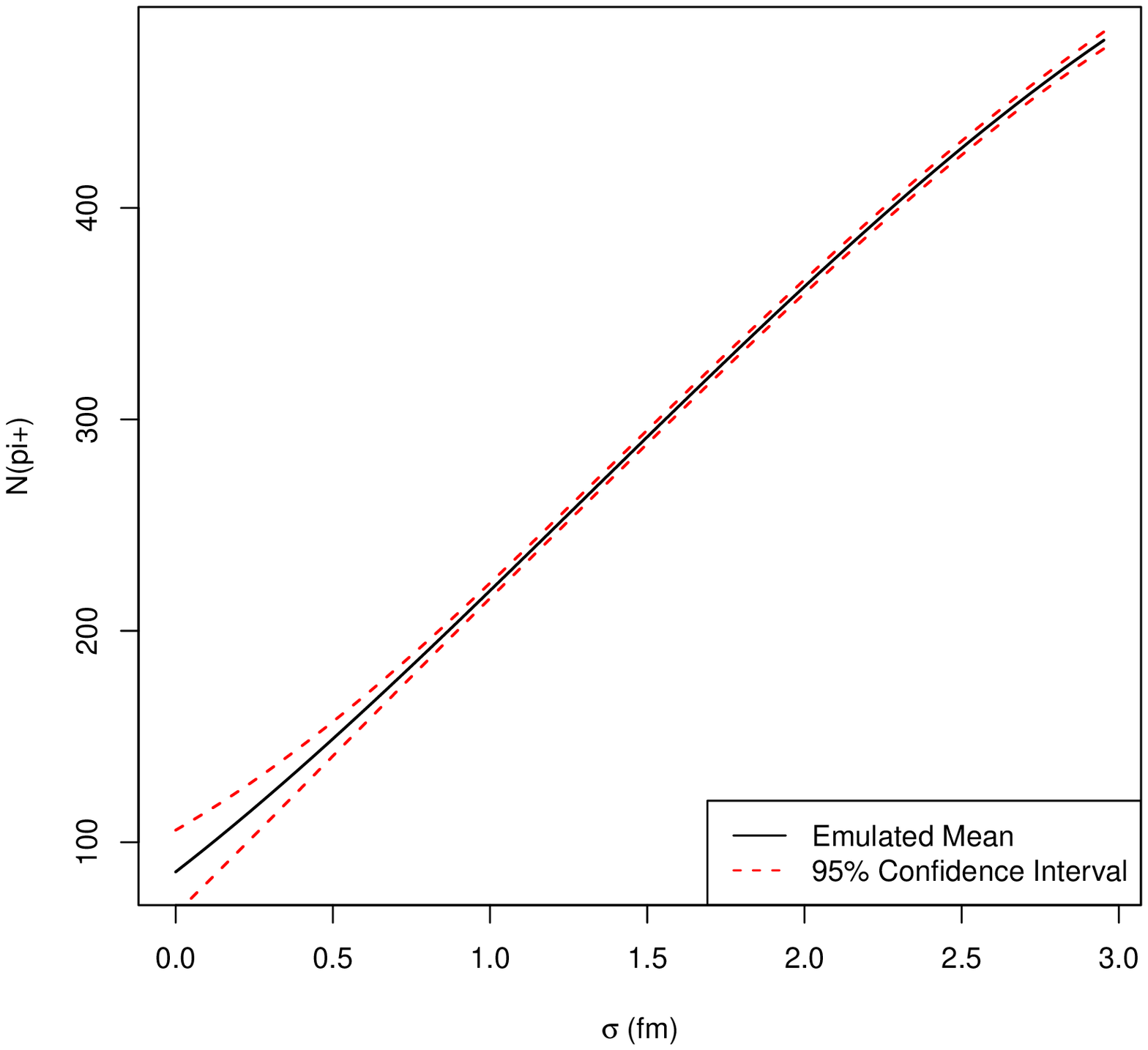}
\includegraphics[width=0.5\textwidth]{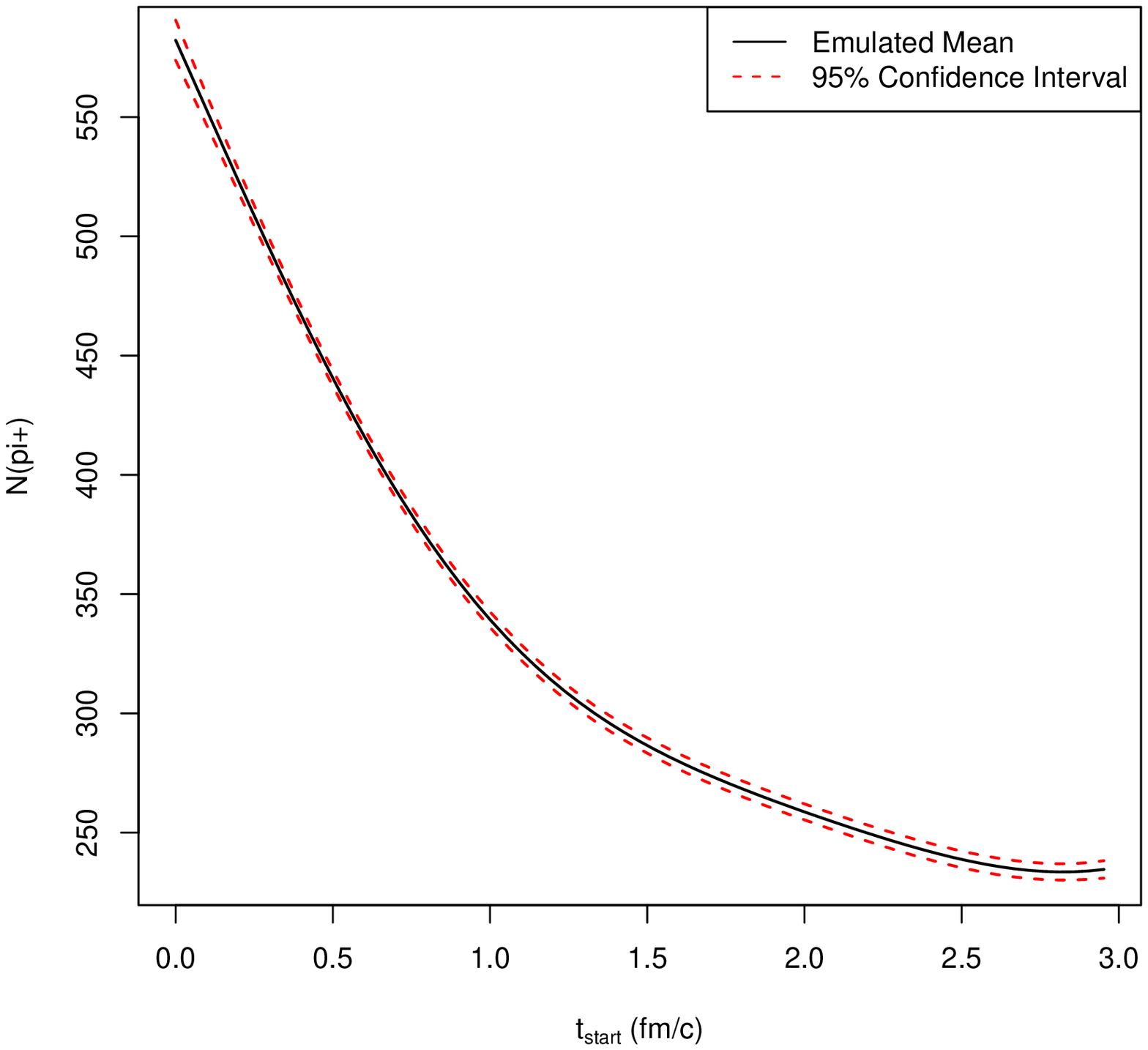}
\caption[Transects at fixed $\sigma$ and $t_{\rm start}$]{Emulated number of
pions at midrapidity
($|y|<0.5$) for central ($b<3.4$ fm) Au+Au collisions at  $\sqrt{s_{\rm
NN}}=200$ GeV. Left: Transect at fixed $t_{\rm start}$ and right: at fixed
$\sigma$.
\label{fig_transects} }
\end{figure}

Fig. \ref{fig_emu_pions} shows the emulated mean number of pions produced in
one collision as a function of the two initial state parameters $t_{\rm start}$
and
$\sigma$. A strong linear correlation between the parameters can be observed.
The
emulator was trained on 30 observations of the model distributed somewhat
uniformly in the 2d parameter space. The variance of the emulator
(Fig. \ref{fig_transects}) in the region where the
simulation data was collected $\sigma \in (0.2, 3.0)$ fm and $t_{\rm start}
\in (0.8,3)$ fm is small, so we can be confident in
the shape of the surface. The variance  increases for $\sigma < 0.8$ fm
which is a region of pure extrapolation, this is the expected behavior.

The linear correlation between start time and kernel width implies that more
smearing in the initial state can be compensated for by prolonged evolution
before the switch to hydrodynamical evolution.

The result for the pion and kaon transverse momentum spectra for the four
parameter combinations are shown in Fig. \ref{fig_dndmt_sig} (right). The
particle yields are now kept constant, but the slope of the spectra is sensitive
to the starting time. The earlier the starting time, the longer the duration
of the hydrodynamic evolution, therefore there is more time to develop larger
radial
flow which leads to flatter transverse mass spectra. The difference in slope
becomes clearly visible at higher $m_T-m_0$ values larger than 1 GeV. 

\begin{figure}[h]
\includegraphics[width=0.5\textwidth]{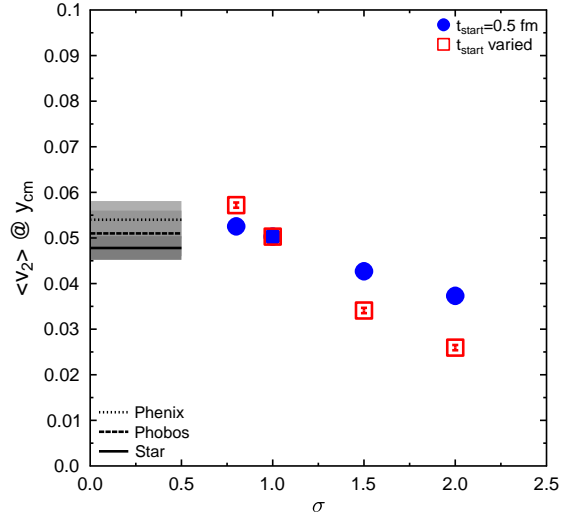}
\centering
\caption[Averaged value of elliptic flow]{The averaged value of elliptic flow of
charged particles at midrapidity ($|y|<0.5$) for mid-central ($b=5-9$ fm) Au+Au
collisions at  $\sqrt{s_{\rm NN}}=200$ GeV as a function of $\sigma$ with fixed
(full circles) and varied $t_{\rm start}$ (open squares) compared to
experimental data represented by black lines (the grey shaded regions indicate
the error bars)\cite{Esumi:2002vy,Manly:2002uq,Ray:2002md}. 
\label{fig_v2_ch_sig} }
\end{figure}

Let us now explore the influence of the different initial state parameters on
the integrated elliptic flow in mid-central ($b=5-9$ fm) Au+Au collisions at
$\sqrt{s_{\rm NN}}=200$ GeV. In Fig. \ref{fig_v2_ch_sig} the averaged value of
the second coefficient of the Fourier decomposition of the azimuthal
distribution of charged particles in momentum space is shown as a function of
the
granularity $\sigma$. The filled circles represent the results for fixed
starting time $t_{\rm start}=0.5$ fm and the open squares show the result for
the identified combinations in Table \ref{tab_sigtstart}. Without changing the
starting time the results for different granularities are compatible to the
experimental data, that is indicated by the lines, even though the particle
numbers are very different. The starting time can be constrained by looking at
the spectra and elliptic flow to be less than 1 fm since later starting times
do not allow for enough flow development during the hydrodynamic evolution.
Just by looking at basic bulk observables like yields, spectra and integrated
elliptic flow, we can constrain the initial state parameters within this
approach to be around $\sigma=1$ fm and $t_{\rm start}=0.5$ fm which we will use
as default in the following Section.

\section{Fluctuating vs. Averaged Initial Conditions}
\label{de_ave}

Even though, we have shown in the last Section that the initial state parameters
can be constrained by bulk observables, this does not necessarily constrain the
granularity of the initial state. Previous studies at lower energies within the
same hybrid approach have shown that there is some insensitivity to the
event-by-event fluctuations of the initial state \cite{Petersen:2010md}. In
contrast to the full event-by-event setup we have discussed so far, one can also
look at calculations from averaged initial conditions. The
default parameters are chosen to generate the initial state and one averages
over 100 initial states from UrQMD to feed a smooth profile in the hydrodynamic
calculation. This provides a different way to tune the initial state granularity
between fluctuating (1 UrQMD event) to averaged smooth initial conditions (100
UrQMD events).   

\begin{figure}[h]
\includegraphics[width=0.5\textwidth]{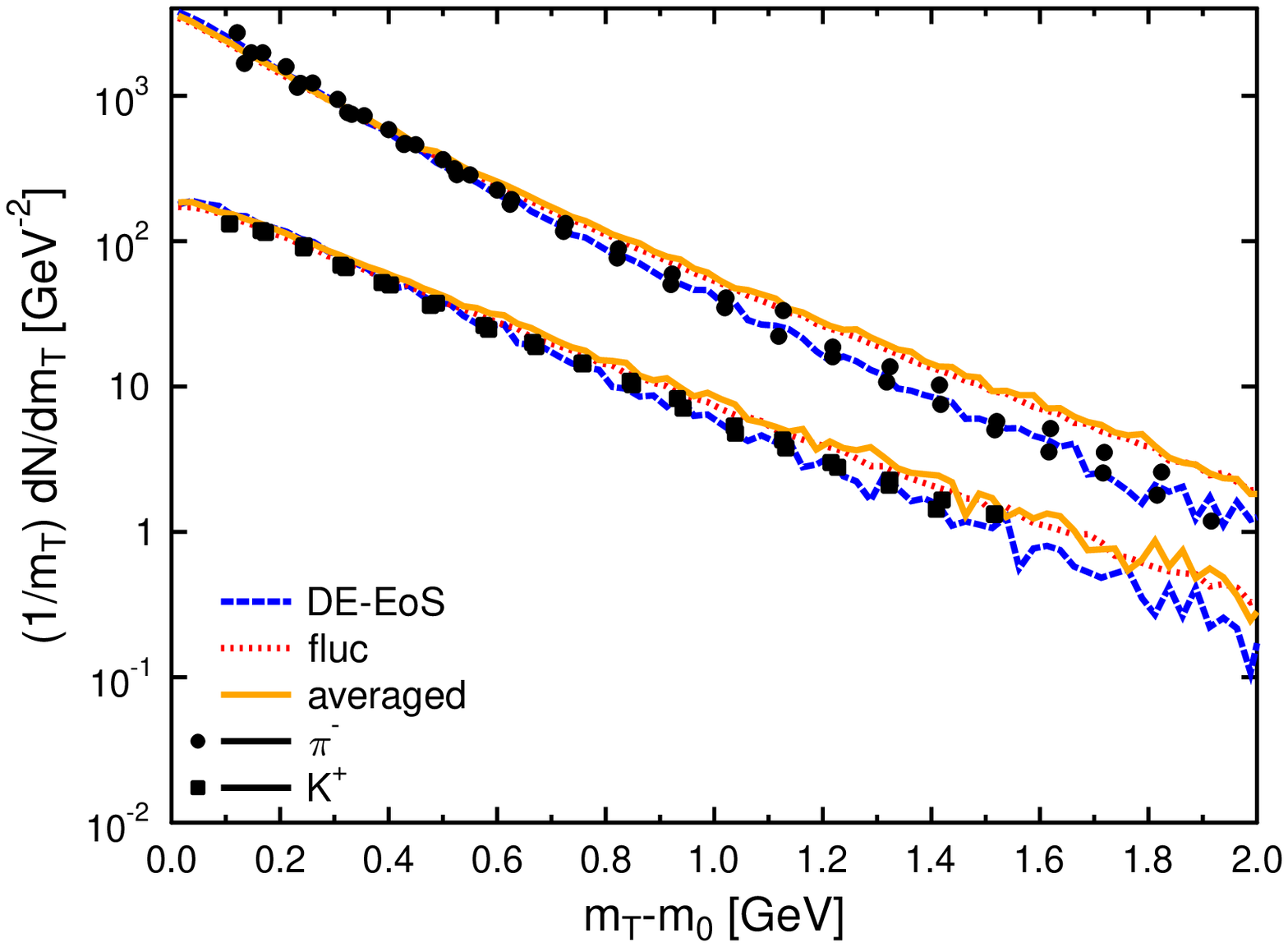}
\includegraphics[width=0.5\textwidth]{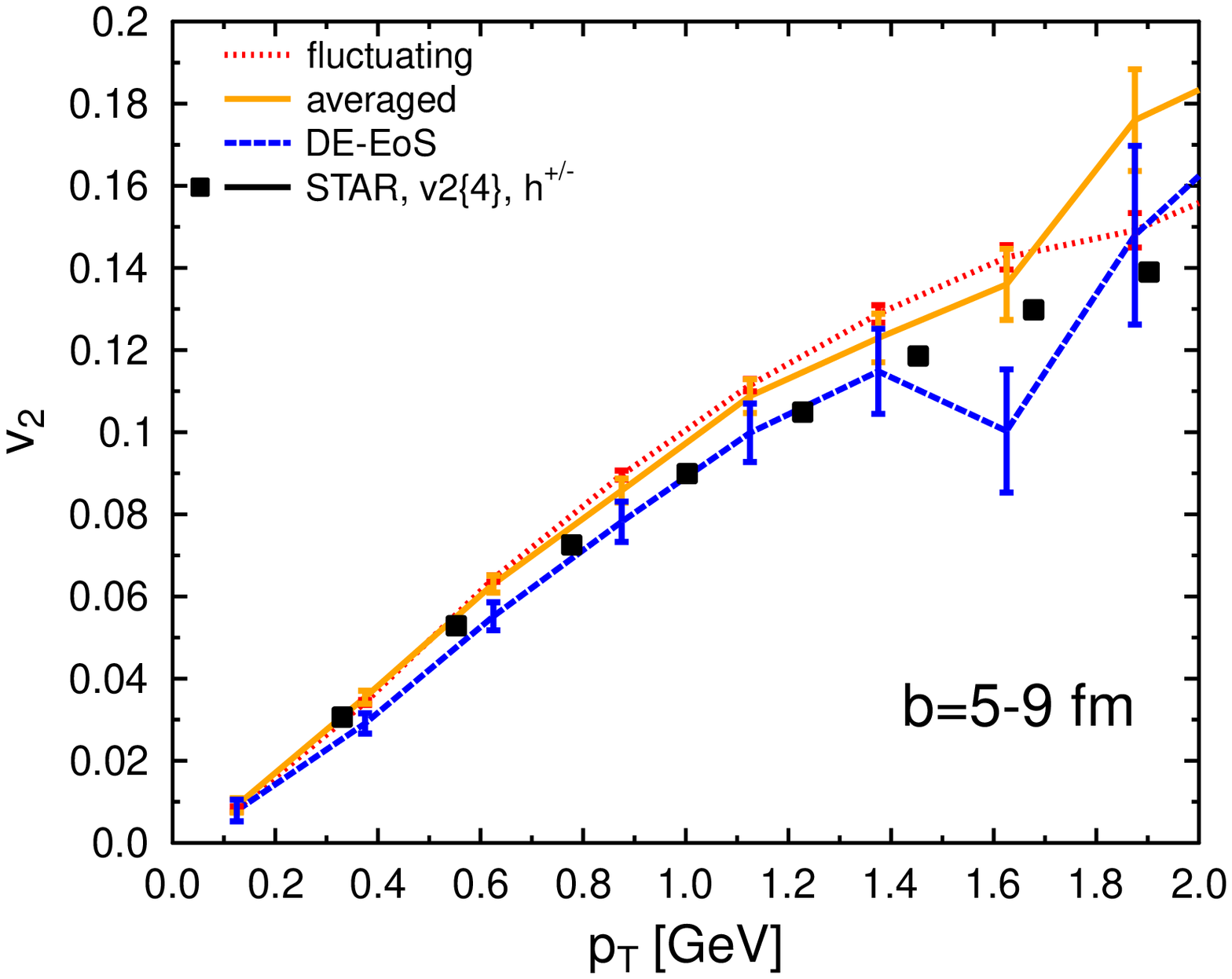}
\caption[Transverse mass spectra and elliptic flow]{Transverse mass spectra for
$\pi^-$ and $K^+$ (left) and elliptic flow as a function of transverse momentum
for pions (right) at midrapidity
($|y|<0.5$) in central/mid-central ($b<3.4$ fm/$b=5-9$ fm) Au+Au collisions at
$\sqrt{s_{\rm NN}}=200$ GeV from the hybrid approach for fluctuating initial
conditions with two different EoS and averaged initial conditions represented by
lines in comparison to experimental data that is shown as symbols
\cite{Adams:2003xp,Adler:2003cb,Arsene:2005mr,:2008ed}. \label{fig_dndmt_flow} }
\end{figure}

In Fig. \ref{fig_dndmt_flow} results for transverse mass spectra (left) and
elliptic flow of pions as a function of transverse momentum (right) are shown
for the fluctuating event-by-event setup and the averaged initial conditions.
The results in this case are not affected since the $v_2$ analysis has been
performed with respect to the reaction plane that is given by the coordinate
system. Using the standard event plane method the elliptic flow results would be
a little higher in the fluctuating case as has been shown in
\cite{Holopainen:2010gz}.
Furthermore, a softer equation of state based on a chiral hadronic Lagrangian
that is coupled to the Polyakov loop including chiral symmetry restoration and a
deconfinement phase transition that reproduces ground state properties and
results from lattice QCD (DE-EoS) \cite{Steinheimer:2009nn,Steinheimer:2009hd,Borsanyi:2010cj} has been applied to give an example of other
'parameters' that may affect the results. As shown in Fig. \ref{fig_dndmt_flow}
(left) the slope of the transverse mass spectra is smaller using the DE-EoS and
leads to a better agreement with experimental data, while the elliptic flow
result (right) is very similar. For the calculations with the other equation of
state the freeze-out transition needs to be adjusted as well and a higher energy
density criterion has been applied. A more detailed study of the freeze-out
procedure is left for a future publication. At this point, it is
mentioned as a further uncertainty. 

\begin{figure}[h]
\includegraphics[width=0.5\textwidth]{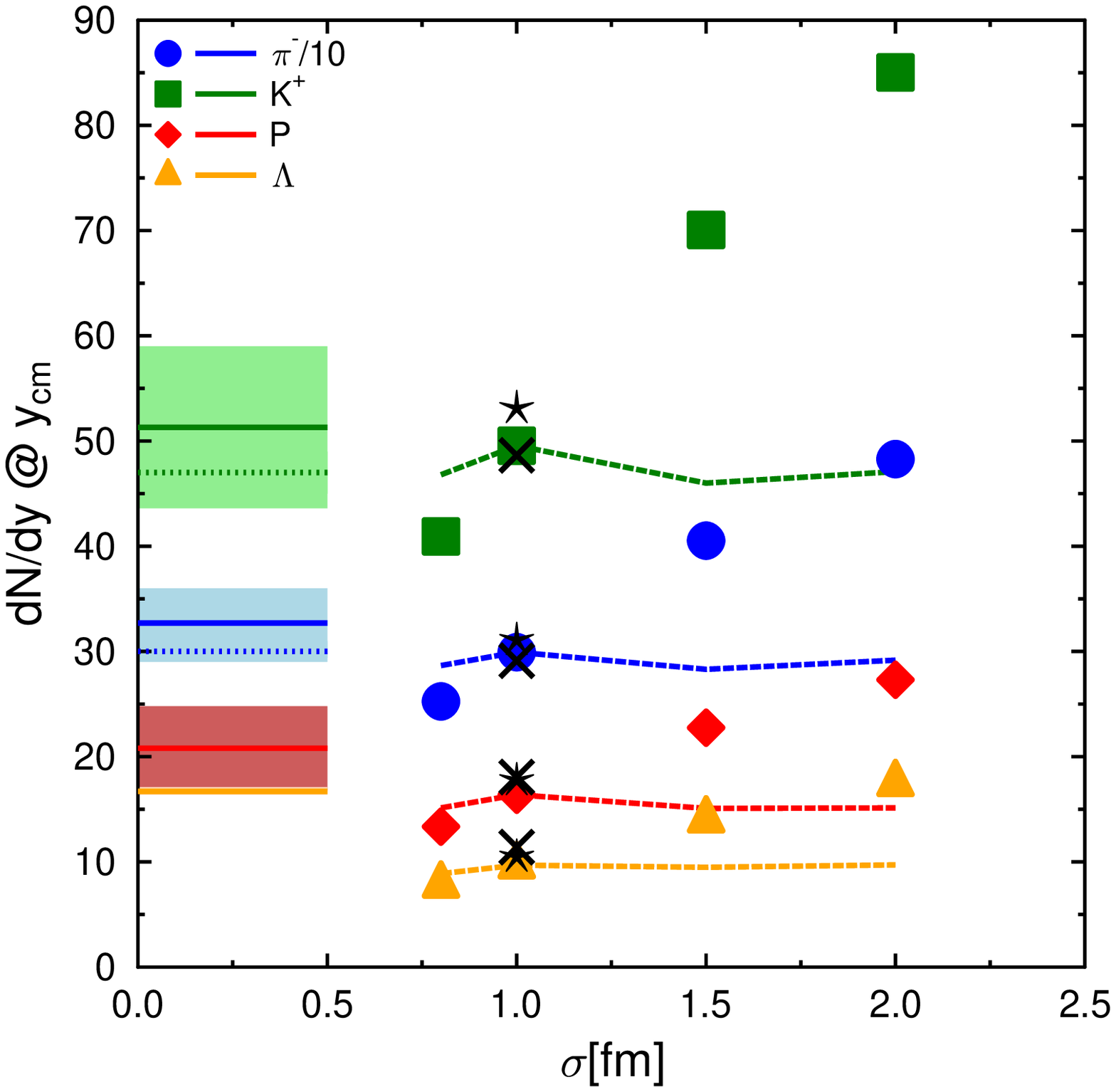}
\includegraphics[width=0.5\textwidth]{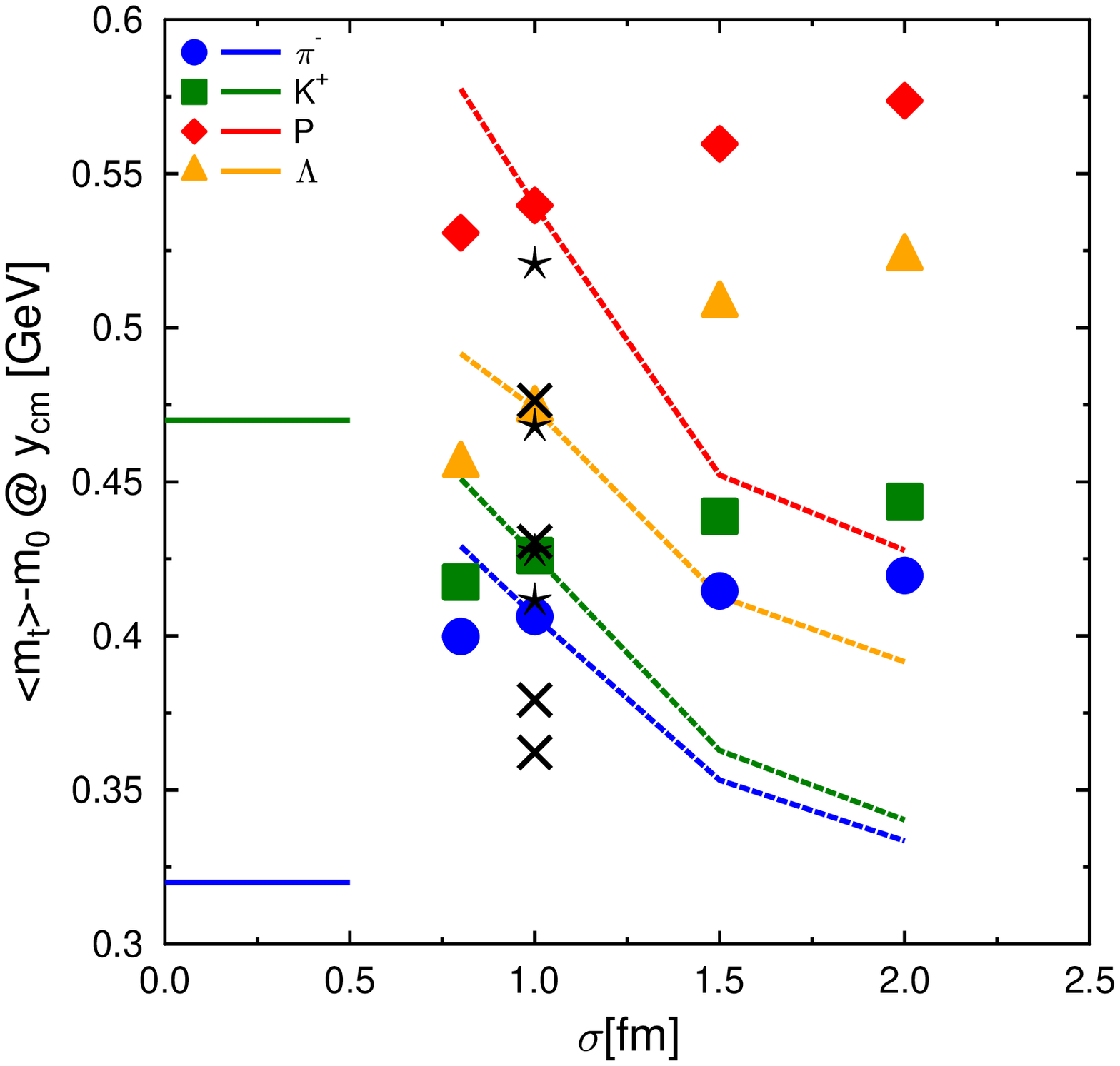}
\caption[Particle Yields and mean transverse mass]{The yields (left) and the
mean transverse mass (right) at midrapidity ($|y|<0.5$) for four different
particle species ($\pi^-$, $K^+$, P, $\Lambda$) are calculated in central
($b<3.4$ fm) Au+Au collisions at $\sqrt{s_{\rm NN}}=200$ GeV. The hybrid
approach for different values of $\sigma$ with a fixed starting time (full
symbols) and a varied starting time (dashed lines) are compared to experimental
data indicated by full (STAR) and dotted (BRAHMS) lines
\cite{Ouerdane:2002gm,Lee:2004bx,Adams:2003xp,Adams:2006ke}. In addition, the
crosses represent the event-by-event calculation with the equation of state
including a deconfinement phase transition (DE-EoS) and results for averaged
initial conditions are shown as stars. \label{fig_mul_mmt} }
\end{figure}

To wrap up the results of the present systematic study on initial state
granularity and bulk observables in heavy ion reactions, in Fig.
\ref{fig_mul_mmt} results for yields and mean transverse momentum of pions,
protons, kaons and $\Lambda$' s are shown for all the 10 different cases that we
have discussed in this paper so far. The full symbols represent the results
where only $\sigma$ has been changed while $t_{\rm start}$ is fixed to be 0.5
fm. The multiplicities are proportional to the smearing parameter, whereas the
radial flow is not as sensitive (with the exception of the $\Lambda$'s). By
varying the starting time according to the result of the emulator (lines), we
confirm that the yields can be kept constant but the mean transverse mass
decreases significantly because of the later transition to the hydrodynamic
evolution. One needs to start early enough to allow to develop radial flow and
elliptic flow. From these two calculations the default parameters for
$\sigma$ and $t_{\rm start}$ are constrained. Still one might vary the
granularity by averaging over initial states instead of final states and the
corresponding results are shown as stars. Bulk observables do not provide a good
handle on the granularity since the results are very similar in the fluctuating
and the averaged case. One needs to calculate fluctuation or correlation
observables to really constrain the amount of initial state fluctuations. The
equation of state including a deconfinement phase transition reduces the
mean transverse masses to achieve a better agreement with the experimental data.
We are not able
to present a complete set of best fit parameters yet, since the freeze-out
procedure needs to be improved before the statistical analysis is extended to
include more parameters and more experimental data sets. 

\section[Summary]{Summary and Conclusion}
\label{sum}

We have explored the capabilities of a novel statistical
analysis utilizing Gaussian process emulators to address
ambiguities in the initial state parameters used in a hybrid approach for the
dynamical evolution of relativistic heavy ion reactions. The smearing parameter
$\sigma$ and the starting time of the hydrodynamic evolution are found
to be correlated. By imposing a constant entropy constraint  one finds 
viable combinations of these two
parameters, which agree well with experimental data. 

The comparison of the fluctuating event-by-event setup to averaged initial
conditions using the same initial state parameters shows that bulk observables
are insensitive to the initial state granularity. This insensitivity
can be exploited in order to
constrain quantities like e.g. the shear viscosity of QCD matter from averaged elliptic
flow results. We note that significant sensitivities to the equation of state remain
which can be utilized to constrain this input parameter. Our
results show that state of the art statistical techniques for executing multi-parameter fits are required to reliably quantify 
properties of interest of hot and dense QCD matter. These techniques
are  now being developed and will lead us to achieve a better understanding of sensitivities on different parameter sets.

\section*{Acknowledgements}
We are grateful to the Open Science Grid for the computing
resources. The authors thank Dirk Rischke for
providing the 1 fluid hydrodynamics code. H.P. acknowledges a Feodor Lynen
fellowship of the Alexander von Humboldt
foundation. This work was supported in part by U.S. department of Energy grant
DE-FG02-05ER41367, NSF grants PHY-09-41373 and DMS-07-57549 and NASA grant NNX09AK60G.


\section*{References}
\begin{thebibliography}{99}
% Save this file and include it in your paper as the bibliography
% or cut and paste directly into your LaTeX

\bibitem{Bass:2000ib}
S.~A.~Bass and A.~Dumitru,
%``Dynamics of hot bulk QCD matter: From the quark-gluon plasma to  hadronic
%freeze-out,''
Phys.\ Rev.\  C {\bf 61}, 064909 (2000)
[arXiv:nucl-th/0001033].
%%CITATION = PHRVA,C61,064909;%%

\bibitem{Teaney:2001av}
D.~Teaney, J.~Lauret and E.~V.~Shuryak,
%``A hydrodynamic description of heavy ion collisions at the SPS and RHIC,''
arXiv:nucl-th/0110037.
%%CITATION = NUCL-TH/0110037;%%

\bibitem{Hirano:2005xf}
T.~Hirano, U.~W.~Heinz, D.~Kharzeev, R.~Lacey and Y.~Nara,
%``Hadronic dissipative effects on elliptic flow in ultrarelativistic
%heavy-ion collisions,''
Phys.\ Lett.\  B {\bf 636}, 299 (2006)
[arXiv:nucl-th/0511046].
%%CITATION = PHLTA,B636,299;%%

\bibitem{Nonaka:2006yn}
C.~Nonaka and S.~A.~Bass,
%``Space-time evolution of bulk QCD matter,''
Phys.\ Rev.\  C {\bf 75}, 014902 (2007)
[arXiv:nucl-th/0607018].
%%CITATION = PHRVA,C75,014902;%%

\bibitem{Petersen:2008dd}
H.~Petersen, J.~Steinheimer, G.~Burau, M.~Bleicher and H.~Stocker,
%``A Fully Integrated Transport Approach to Heavy Ion Reactions with an
%Intermediate Hydrodynamic Stage,''
Phys.\ Rev.\  C {\bf 78}, 044901 (2008)
[arXiv:0806.1695 [nucl-th]].
%%CITATION = PHRVA,C78,044901;%%

\bibitem{Werner:2010aa}
K.~Werner, I.~Karpenko, T.~Pierog, M.~Bleicher and K.~Mikhailov,
%``Event-by-Event Simulation of the Three-Dimensional Hydrodynamic Evolution
%from Flux Tube Initial Conditions in Ultrarelativistic Heavy Ion
%Collisions,''
Phys.\ Rev.\  C {\bf 82}, 044904 (2010)
[arXiv:1004.0805 [nucl-th]].
%%CITATION = PHRVA,C82,044904;%%

\bibitem{Agakishiev:2010ur}
H.~Agakishiev {\it et al.},
%``Measurements of Dihadron Correlations Relative to the Event Plane in Au+Au
%Collisions at $\sqrt{s_{NN}}=200$ GeV,''
arXiv:1010.0690 [nucl-ex].
%%CITATION = ARXIV:1010.0690;%%

\bibitem{Alver:2010dn}
B.~H.~Alver, C.~Gombeaud, M.~Luzum and J.~Y.~Ollitrault,
%``Triangular flow in hydrodynamics and transport theory,''
Phys.\ Rev.\  C {\bf 82}, 034913 (2010)
[arXiv:1007.5469 [nucl-th]].
%%CITATION = PHRVA,C82,034913;%%

\bibitem{Schenke:2010rr}
B.~Schenke, S.~Jeon and C.~Gale,
%``Elliptic and triangular flow in event-by-event (3+1)D viscous
%hydrodynamics,''
arXiv:1009.3244 [hep-ph].
%%CITATION = ARXIV:1009.3244;%%

\bibitem{Petersen:2010cw}
H.~Petersen, G.~Y.~Qin, S.~A.~Bass and B.~Muller,
%``Triangular flow in event-by-event ideal hydrodynamics in Au+Au collisions
%at $\sqrt{s_{\rm NN}}=200A$ GeV,''
Phys.\ Rev.\  C {\bf 82}, 041901 (2010)
[arXiv:1008.0625 [nucl-th]].
%%CITATION = PHRVA,C82,041901;%%

\bibitem{Qin:2010pf}
G.~Y.~Qin, H.~Petersen, S.~A.~Bass and B.~Muller,
%``Translation of collision geometry fluctuations into momentum anisotropies
%in relativistic heavy-ion collisions,''
arXiv:1009.1847 [nucl-th].
%%CITATION = ARXIV:1009.1847;%%

\bibitem{Mocsy:2010um}
A.~Mocsy and P.~Sorensen,
%``The Sound of the Little Bangs,''
arXiv:1008.3381 [hep-ph].
%%CITATION = ARXIV:1008.3381;%%

\bibitem{Bass:1998ca}
S.~A.~Bass {\it et al.},
%``Microscopic models for ultrarelativistic heavy ion collisions,''
Prog.\ Part.\ Nucl.\ Phys.\  {\bf 41}, 255 (1998)
[Prog.\ Part.\ Nucl.\ Phys.\  {\bf 41}, 225 (1998)]
[arXiv:nucl-th/9803035].
%%CITATION = PPNPD,41,225;%%

\bibitem{Bleicher:1999xi}
M.~Bleicher {\it et al.},
%``Relativistic hadron hadron collisions in the ultra-relativistic quantum
%molecular dynamics model,''
J.\ Phys.\ G {\bf 25}, 1859 (1999)
[arXiv:hep-ph/9909407].
%%CITATION = JPHGB,G25,1859;%%

\bibitem{NilssonAlmqvist:1986rx}
B.~Nilsson-Almqvist and E.~Stenlund,
%``Interactions Between Hadrons And Nuclei: The Lund Monte Carlo, Fritiof
%Version 1.6,''
Comput.\ Phys.\ Commun.\  {\bf 43}, 387 (1987).
%%CITATION = CPHCB,43,387;%%

\bibitem{Sjostrand:1993yb}
T.~Sjostrand,
%``High-energy physics event generation with PYTHIA 5.7 and JETSET 7.4,''
Comput.\ Phys.\ Commun.\  {\bf 82}, 74 (1994).
%%CITATION = CPHCB,82,74;%%

\bibitem{Bleicher:1998wd}
M.~Bleicher {\it et al.},
%``Fluctuations and inhomogenities of energy density and isospin in  Pb + Pb
%at the SPS,''
Nucl.\ Phys.\  A {\bf 638}, 391 (1998).
%%CITATION = NUPHA,A638,391;%%

\bibitem{Grassi:2005pm}
F.~Grassi, Y.~Hama, O.~Socolowski and T.~Kodama,
%``Results on transverse mass spectra obtained with NeXSPheRIO,''
J.\ Phys.\ G {\bf 31}, S1041 (2005).
%%CITATION = JPHGB,G31,S1041;%%

\bibitem{Andrade:2005tx}
R.~Andrade, F.~Grassi, Y.~Hama, T.~Kodama, O.~.~J.~Socolowski and B.~Tavares,
%``NeXSPheRIO results on elliptic flow at RHIC and connection with
%thermalization,''
Eur.\ Phys.\ J.\  A {\bf 29}, 23 (2006)
[arXiv:nucl-th/0511021].
%%CITATION = EPHJA,A29,23;%%

\bibitem{Andrade:2006yh}
R.~Andrade, F.~Grassi, Y.~Hama, T.~Kodama and O.~.~J.~Socolowski,
%``On the necessity to include event-by-event fluctuations in experimental
%evaluation of elliptical flow,''
Phys.\ Rev.\ Lett.\  {\bf 97}, 202302 (2006)
[arXiv:nucl-th/0608067].
%%CITATION = PRLTA,97,202302;%%

\bibitem{Tavares:2007mu}
B.~M.~Tavares, H.~J.~Drescher and T.~Kodama,
%``Effects of nucleus initialization on event-by-event observables,''
Braz.\ J.\ Phys.\  {\bf 37}, 41 (2007)
[arXiv:hep-ph/0702224].
%%CITATION = BJPHE,37,41;%%

\bibitem{Andrade:2008xh}
R.~P.~G.~Andrade, F.~Grassi, Y.~Hama, T.~Kodama and W.~L.~Qian,
%``Importance of Granular Structure in the Initial Conditions for the Elliptic
%Flow,''
Phys.\ Rev.\ Lett.\  {\bf 101}, 112301 (2008)
[arXiv:0805.0018 [hep-ph]].
%%CITATION = PRLTA,101,112301;%%

\bibitem{Petersen:2010di}
H.~Petersen, T.~Renk and S.~A.~Bass,
%``Medium-modified Jets and Initial State Fluctuations as Sources of Charge
%Correlations Measured at RHIC,''
arXiv:1008.3846 [nucl-th].
%%CITATION = ARXIV:1008.3846;%%

\bibitem{Rischke:1995ir}
D.~H.~Rischke, S.~Bernard and J.~A.~Maruhn,
%``Relativistic hydrodynamics for heavy ion collisions. 1. General aspects and
%expansion into vacuum,''
Nucl.\ Phys.\  A {\bf 595}, 346 (1995)
[arXiv:nucl-th/9504018].
%%CITATION = NUPHA,A595,346;%%

\bibitem{Rischke:1995mt}
D.~H.~Rischke, Y.~Pursun and J.~A.~Maruhn,
%``Relativistic hydrodynamics for heavy ion collisions. 2. Compression of
%nuclear matter and the phase transition to the quark - gluon plasma,''
Nucl.\ Phys.\  A {\bf 595}, 383 (1995)
[Erratum-ibid.\  A {\bf 596}, 717 (1996)]
[arXiv:nucl-th/9504021].
%%CITATION = NUPHA,A595,383;%%

\bibitem{Petersen:2009mz}
H.~Petersen, J.~Steinheimer, M.~Bleicher and H.~Stocker,
%``$<m_T>$ excitation function: Freeze-out and equation of state dependence,''
J.\ Phys.\ G {\bf 36}, 055104 (2009)
[arXiv:0902.4866 [nucl-th]].
%%CITATION = JPHGB,G36,055104;%%

% No SPIRES record found for cite request 1198/01621450

\bibitem{OHag:2006}
Anthony O'Hagan,
% Bayesian analysis of computer code outputs: A tutorial.
Reliability Engineering \& System Safety, 91 (10-11):1290--1300, (2006), The
Fourth International Conference on Sensitivity Analysis of Model Output (SAMO
2004) - SAMO 2004.
%         DOI = "10.1016/j.ress.2005.11.025",
%http://www.sciencedirect.com/science/article/B6V4T-4J0WRB5-5/2/ca62ab87249507e8
%ba221ab8b419f77c",

\bibitem{Oakl:OHag:2002}
Jeremy E. Oakley and Anthony O'Hagan,
%"Bayesian inference for the uncertainty distribution of computer model outputs",
Biometrika, {\bf 89 (4)} :169-184, (2002). 
%         DOI = "10.1093/biomet/89.4.769",
%         URL = "http://biomet.oxfordjournals.org/content/89/4/769.abstract",
%      Eprint =
%"http://biomet.oxfordjournals.org/content/89/4/769.full.pdf+html",


\bibitem{Oakl:OHag:2004}
Jeremy E. Oakley and Anthony O'Hagan,
% "Probabilistic sensitivity analysis of complex models: a
%                  {B}ayesian approach",
J.\ Roy.\ Stat.\ Soc. B,{\bf 66},(2004).  
%         DOI = "10.1111/j.1467-9868.2004.05304.x",

\bibitem{Baya:Berg:Paul:etal:2007}
 M. J. Bayarri, James O. Berger, Rui Paulo, Jerome Sacks, John A. Cafeo, James C. Cavendish, Chin-Hsu Lin and Jian Tu,
%       Title = "A Framework for Validation of Computer Models",
Technometrics, {\bf 49 (2)}:138-154, (2007).        

\bibitem{Higd:Gatt:etal:2008}
David Higdon, James Gattiker, Brian Williams and Maria Rightley,
%       Title = "Computer Model Calibration using High-Dimensional Output",
J.\ Am.\ Stat.\ Assoc., {\bf 103 (482)}:570-583,(2008).
%         DOI = "10.1198/016214507000000888",
%         URL = "http://pubs.amstat.org/doi/abs/10.1198/016214507000000888",
%      Eprint = "http://pubs.amstat.org/doi/pdf/10.1198/016214507000000888",

\bibitem{Chil:Delf:1999}
Jean-Paul Chil{\`e}s and Pierre Delfiner,
%       Title = "Geostatistics: Modeling Spatial Uncertainty",
John Wiley \& Sons, New York, NY, (1999).  

%        ISBN = "0-471-08315-1",

\bibitem{Cres:1993}
Noel A. C. Cressie,
%       Title = "Statistics for Spatial Data",
John Wiley \& Sons, New York, NY, (1993).
%        ISBN = "0-471-00255-0",


\bibitem{Kenn:OHag:2000} 
Marc C. Kennedy and Anthony O'Hagan,
%       Title = "Predicting the output from a complex computer code when
%                  fast approximations are available",
Biometrika, {\bf 87(1)}:1-13, (2000). 
%         DOI = "10.1093/biomet/87.1.1",


\bibitem{Sack:Welc:Mitc:Wynn:1989}
Jerome Sacks, William J. Welch, Toby J. Mitchell and Henry P. Wynn,
%       Title = "Design and Analysis of Computer Experiments",
Stat.\ Sci., {\bf 4 (4)}:409-435, November (1989). 


\bibitem{Sant:Will:Notz:2003}
Thomas J. Santner, Brian J. Williams and William Notz,
%       Title = "The Design and Analysis of Computer Experiments",
Springer Series in Statistics, Springer-Verlag, New York, NY, (2003).
%      Mynoee = "P 87 or so, Section 4.1.2, has full Bayesian calcs",
%   Publisher = sv,
%     Address = "New York, NY",
%        Year = 2003,
%}



\bibitem{Adams:2003xp}
J.~Adams {\it et al.}  [STAR Collaboration],
%``Identified particle distributions in p p and Au + Au collisions at
%s**(1/2) = 200-GeV,''
Phys.\ Rev.\ Lett.\  {\bf 92}, 112301 (2004)
[arXiv:nucl-ex/0310004].
%%CITATION = PRLTA,92,112301;%%

\bibitem{Adler:2003cb}
S.~S.~Adler {\it et al.}  [PHENIX Collaboration],
%``Identified charged particle spectra and yields in Au + Au collisions at
%s(NN)**(1/2) = 200-GeV,''
Phys.\ Rev.\  C {\bf 69}, 034909 (2004)
[arXiv:nucl-ex/0307022].
%%CITATION = PHRVA,C69,034909;%%

\bibitem{Arsene:2005mr}
I.~Arsene {\it et al.}  [BRAHMS Collaboration],
%``Centrality dependent particle production at y = 0 and y approx. 1 in Au  +
%Au collisions at s(NN)**(1/2) = 200-GeV,''
Phys.\ Rev.\  C {\bf 72}, 014908 (2005)
[arXiv:nucl-ex/0503010].
%%CITATION = PHRVA,C72,014908;%%

\bibitem{Esumi:2002vy}
S.~Esumi  [PHENIX Collaboration],
%``Identified charged particle azimuthal anisotropy in PHENIX at RHIC,''
Nucl.\ Phys.\  A {\bf 715}, 599 (2003)
[arXiv:nucl-ex/0210012].
%%CITATION = NUPHA,A715,599;%%

\bibitem{Manly:2002uq}
S.~Manly {\it et al.}  [PHOBOS Collaboration],
%``Flow and Bose-Einstein correlations in Au Au collisions at RHIC,''
Nucl.\ Phys.\  A {\bf 715}, 611 (2003)
[arXiv:nucl-ex/0210036].
%%CITATION = NUPHA,A715,611;%%

\bibitem{Ray:2002md}
R.~L.~Ray  [STAR Collaboration],
%``Correlations, fluctuations, and flow measurements from the STAR
%experiment,''
Nucl.\ Phys.\  A {\bf 715}, 45 (2003)
[arXiv:nucl-ex/0211030].
%%CITATION = NUPHA,A715,45;%%

\bibitem{Petersen:2010md}
H.~Petersen and M.~Bleicher,
%``Eccentricity fluctuations in an integrated hybrid approach: Influence on
%elliptic flow,''
Phys.\ Rev.\  C {\bf 81}, 044906 (2010)
[arXiv:1002.1003 [nucl-th]].
%%CITATION = PHRVA,C81,044906;%%

\bibitem{:2008ed}
B.~I.~Abelev {\it et al.}  [STAR Collaboration],
%``Centrality dependence of charged hadron and strange hadron elliptic flow
%from sqrt(s_NN) = 200 GeV Au+Au collisions,''
Phys.\ Rev.\  C {\bf 77}, 054901 (2008)
[arXiv:0801.3466 [nucl-ex]].
%%CITATION = PHRVA,C77,054901;%%

\bibitem{Holopainen:2010gz}
H.~Holopainen, H.~Niemi and K.~J.~Eskola,
%``Event-by-event hydrodynamics and elliptic flow from fluctuating initial
%state,''
arXiv:1007.0368 [hep-ph].
%%CITATION = ARXIV:1007.0368;%%

\bibitem{Steinheimer:2009nn}
  J.~Steinheimer, V.~Dexheimer, H.~Petersen, M.~Bleicher, S.~Schramm and H.~Stoecker,
  %``Hydrodynamics with a chiral hadronic equation of state including quark
  %degrees of freedom,''
  Phys.\ Rev.\  C {\bf 81}, 044913 (2010)
  [arXiv:0905.3099 [hep-ph]].
  %%CITATION = PHRVA,C81,044913;%%

\bibitem{Steinheimer:2009hd}
  J.~Steinheimer, S.~Schramm and H.~Stocker,
  %``An effective chiral Hadron-Quark Equation of State Part I: Zero
  %baryochemical potential,''
  arXiv:0909.4421 [hep-ph].
  %%CITATION = ARXIV:0909.4421;%%
%\cite{Borsanyi:2010cj}
\bibitem{Borsanyi:2010cj}
  S.~Borsanyi, G.~Endrodi, Z.~Fodor {\it et al.},
  %``The QCD equation of state with dynamical quarks,''
  JHEP {\bf 1011 } (2010)  077
  [arXiv:1007.2580 [hep-lat]].

\bibitem{Ouerdane:2002gm}
D.~Ouerdane  [BRAHMS Collaboration],
%``Rapidity dependence of charged particle yields for Au + Au at s(NN)**(1/2)
%= 200-GeV,''
Nucl.\ Phys.\  A {\bf 715}, 478 (2003)
[arXiv:nucl-ex/0212001].
%%CITATION = NUPHA,A715,478;%%

\bibitem{Lee:2004bx}
J.~H.~Lee {\it et al.}  [BRAHMS Collaboration],
%``Rapidity dependent strangeness measurements in BRAHMS experiment at RHIC,''
J.\ Phys.\ G {\bf 30}, S85 (2004).
%%CITATION = JPHGB,G30,S85;%%

\bibitem{Adams:2006ke}
J.~Adams {\it et al.}  [STAR Collaboration],
%``Scaling Properties of Hyperon Production in Au+Au Collisions at
%sqrt(s_NN) = 200 GeV,''
Phys.\ Rev.\ Lett.\  {\bf 98}, 062301 (2007)
[arXiv:nucl-ex/0606014].
%%CITATION = PRLTA,98,062301;%%


\end{thebibliography}
\end{document}